%% file: ssorsfHEP.tex
\documentclass{elsart}
\usepackage{macros_ALPHA}
\usepackage[dvips]{epsfig}
\begin{document}
\input title.tex
\input sect1.tex

\input sect2.tex
\input sect3.tex
\bibliography{lattice_ALPHA}      
\bibliographystyle{h-elsevier}    
\end{document}

%% file: title.tex
\begin{frontmatter}

\begin{flushright}
DESY 99-155 \\
ROM2F 99-37 \\
hep-lat/9910024
\end{flushright}

\title{
SSOR preconditioning in simulations of \\
the QCD Schr\"odinger functional
}
\vbox{
\centerline{
\epsfxsize=2.5 true cm
\epsfbox{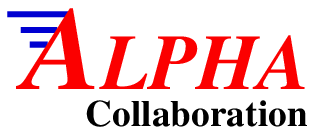}}
}
\author[INFN]{Marco Guagnelli} and
\author[DESY]{Jochen Heitger}
\address[INFN]{
Dipartimento di Fisica, Universit\`{a} di Roma ``Tor Vergata'' \\
and INFN, Sezione di Roma II \\
Via della Ricerca Scientifica 1, I-00133 Rome, Italy
}
\address[DESY]{
Deutsches Elektronen-Synchrotron, DESY Zeuthen \\
Platanenallee 6, D-15738 Zeuthen, Germany
}

\maketitle
\begin{abstract}

We report on a parallelized implementation of SSOR preconditioning for
$\Or(a)$ improved lattice QCD with Schr\"odinger functional boundary
conditions.
Numerical simulations in the quenched approximation at parameters in the
light quark mass region demonstrate that a performance gain of a factor 
$\sim$ 1.5 over even-odd preconditioning can be achieved.

\end{abstract}

\keyword{
lattice QCD; $\Or(a)$ improvement; Schr\"odinger functional; 
SSOR pre\-con\-di\-tioning; parallelization; quenched simulations}\\
\PACS{11.15.Ha; 12.38.Gc; 2.60.Cb; 2.60.Dc}
\endkeyword

\end{frontmatter}
\cleardoublepage

%% file: sect1.tex
\section{Introduction}
\label{Intro}
One of the severe problems in lattice QCD from the practical point of
view is the numerical inversion of sparse matrices.
It nearly enters every Monte Carlo simulation, either in the quenched
approximation of the theory in applying the inverse of the discretized
Dirac operator on source vectors for the computation of quark
propagators, or in full QCD with Hybrid Monte Carlo like algorithms,
where a similar operation is necessary when calculating the fermionic
force in order to include the quark field dynamics in the updating
process of the gauge fields.

In recent years substantial progress has been made by the use of Krylov
subspace iterative solvers in conjunction with preconditioning
techniques.  
(See e.g.~\cite{matrix:GL} for a textbook reference and the reviews
\cite{solver:review,solver:templ,solver:frommer_lat96}, which contain
an extensive bibliography.)
Popular choices for the inverter to be mentioned in the context of
lattice simulations are the Conjugate Gradient (CG) algorithm
\cite{solver:cg1,solver:cg2}, the Minimal Residual (MR) algorithm
\cite{solver:mr1,solver:mr2} and above all the stabilized Bi-Conjugate
Gradient (BiCGStab) algorithm \cite{solver:bicg_vorst,solver:bicg_gutkn}.
The latter now seems commonly established as the most efficient solver in
a vast majority of QCD applications
\cite{solver:bicg_wupp,solver:bicg_andreas}, particularly in the
parameter region of small quark masses.

In the area of preconditioning, any new (and potentially competitive)
idea should be confronted with an even-odd (e/o) decomposition of the
Dirac matrix \cite{eo:degrand1,eo:degrand2}, which both for ordinary
Wilson fermions and together with an $\Or(a)$ improvement term
\cite{jansen:1997yt} has become the state-of-the-art method.
An earlier step towards an alternative approach was the incomplete LU
factorization \cite{ilu:oyanagi} utilizing a 
\emph{globally}-lexicographic ordering of the lattice points and thereby
already related to SSOR.
However, it turns out to be unsatisfactory when an implementation on
grid-oriented parallel computers is envisaged \cite{ilu:hockney_lat89}.
Finally, the Wuppertal group invented a parallelization of symmetric
successive overrelaxed (SSOR) preconditioning, a variant of the
classical Gauss-Seidel iteration \cite{matrix:GL}, resting upon a
\emph{locally}-lexicographic ordering of the points within a given
(sub-)grid of the total space-time lattice
\cite{ssor:color_wupp,ssor:wils_wupp,ssor:impr_wupp_lat97,ssor:impr_wupp,ssor:impr_wupp_lat98}.
They showed that at least for Wilson fermions an SSOR preconditioner
can perform much better that the standard e/o one.

Since many comparative studies of the properties of the various
inversion algorithms and preconditioning methods are already available
in the literature
\cite{solver:frommer_lat96,solver:bicg_andreas,ilu:hockney_lat89},
we restrict ourselves in the sequel exclusively to the
case of the $\Or(a)$ improved Wilson-Dirac operator involving the
Sheikholeslami-Wohlert clover term \cite{impr:SW} within the
Schr\"odinger functional of QCD \cite{SF:LNWW,schlad:rainer}.

The \emph{Schr\"odinger functional (SF)} is defined as the partition
function of QCD in a space-time cylinder of extension $L^3\times T$ with
periodic boundary conditions in the space directions and (inhomogeneous)
Dirichlet boundary conditions at times $0$ and $T$.
As detailed e.g.~in refs.~\cite{schlad:rainer,alpha:SU3}, the SF has
proven to be a valuable tool for computing the running coupling constant
in a finite-size scaling analysis as well as for extracting
phenomenological quantities from simulations in physically large volumes
\cite{msbar:pap2,msbar:pap3}.
On the other hand, the \emph{$\Or(a)$ improved Wilson-Dirac operator} is
now a good candidate to probe continuum QCD by means of numerical
simulations on the lattice:
most coefficients multiplying the counterterms required for a complete
removal of the leading $\Or(a)$ lattice artifacts of action and quark
currents are known non-perturbatively in the quenched approximation
\cite{schlad:rainer,impr:pap3,tsuk97:rainer}, and also for the action in
the case of two flavours of dynamical quarks \cite{impr:csw_nf2}.
Therefore, a quantitative investigation of the performance of a 
(parallelized) SSOR preconditioner for the SF setup incorporating
$\Or(a)$ improvement appears to be of natural interest.
This is what we intend with the present communication.

One might ask what should be different from periodic boundary conditions
in space \emph{and} time.
At first, it is in principle not excluded that with Dirichlet boundary
conditions a (specifically preconditioned) solver has generally lower
iteration numbers.
Another point concerns a definite advantage in the actual implementation, 
since the SF allows to circumvent inefficient conditional statements
as will be explained later.

%% file: sect2.tex
\section{SSOR preconditioning}
\label{SSOR}
The basic problem posed is to solve a system of linear equations of the
type
\be
\cl{A}X=Y\quad\Leftrightarrow\quad
R\equiv\cl{A}X-Y=0\,.
\label{syst}
\ee
To fix some notation, small Greek letters denote scalars, capital Latin
ones vectors with components in small letters, and matrices have 
calligraphic symbols; $(X,Y)=\sum_i x_i^*y_i$ is the usual scalar product.
In lattice QCD, $\cl{A}$ represents the discretized Dirac operator, a huge
sparse matrix of rank $\Omega\times4\times3$ emerging from the nearest
neighbour couplings of the quark and gluon field variables in position
space after discretization of the interacting continuum theory.
More precisely, $\Omega$ is the volume of the four-dimensional space-time
lattice, and the latter factors are connected to Dirac and SU(3) colour
spaces.
Physically, the solution of eq.~(\ref{syst}) can be regarded as a
fermionic Green function (i.e.~a quark propagator).

SSOR preconditioning consists of solving the system
\be
\cl{V}_1^{-1}\cl{A}\cl{V}_2^{-1}\tilde{X}=\tilde{Y}\,,\quad
\tilde{X}=\cl{V}_2X\,,\quad \tilde{Y}=\cl{V}_1^{-1}Y
\label{ssor}
\ee
instead of $\cl{A}X=Y$, where in the present context $\cl{A}$ stands for
$\wdo$, the lattice Dirac operator of Wilson fermions supplemented with a
local $\Or(a)$ improvement term $a\smunu\pl{F}{\mu\nu}$, which is composed
of the SU(3)--valued gauge link variables $\gf{U}{\mu}$ to form a clover
leaf \cite{impr:SW}:
\bea
\wdo
& \,=\, &
{{\bf 1}}\delta_{x,y}\nonumber\\
&       &
-\,\,\kappa\sum_{\mu}\Big\{
({\bf{1}}-\gmu)\gf{U}{\mu}\delta_{x,y-a\hat{\mu}}
+({\bf{1}}+\gmu)\gfdn{U}{\mu}{\mu}^+\delta_{x,y+a\hat{\mu}}
\Big\}\nonumber\\
&       &
+\,\,\csw\,\frac{i}{2}\,a\,\kappa
\sum_{\mu,\nu}\smunu\pl{F}{\mu\nu}\,\delta_{x,y}
\label{w_dir_clover}
\eea
with $\pl{F}{\mu\nu}$ equal to
\bea
\frac{1}{8}\,\bigg\{\,
&   &
\Big[\,\gf{U}{\mu}\gfup{U}{\mu}{\nu}\gfup{U}{\nu}{\mu}^+\gf{U}{\nu}^+
\nonumber\\
&   &
+\,\,\gf{U}{\nu}\gfud{U}{\nu}{\mu}{\mu}^+
\gfdn{U}{\mu}{\nu}^+\gfdn{U}{\mu}{\mu}\nonumber\\
&   &
+\,\,\gfdn{U}{\mu}{\mu}^+\gfdd{U}{\nu}{\mu}{\nu}^+\gfdd{U}{\nu}{\mu}{\mu}
\gfdn{U}{\nu}{\nu}\nonumber\\
&   &
+\,\,\gfdn{U}{\nu}{\nu}^+\gfdn{U}{\nu}{\mu}\gfdu{U}{\nu}{\mu}{\nu}
\gf{U}{\mu}^+\,\Big]\nonumber\\
&   &
-\,\,\Big[\,\cdots\,\Big]^+\,\,\,\bigg\}\,.
\label{clover}
\eea
$\csw$ is an improvement coefficient non-perturbatively determined 
in quenched and two-flavour QCD \cite{impr:pap3,impr:csw_nf2}.
(The slight modification to (\ref{w_dir_clover}) induced by the Dirichlet
boundary conditions in time direction are not written out here.)
If one introduces the decomposition\footnote{
The case of Wilson-Dirac fermions is recovered by setting $\calD$ to the
unit matrix.}
\be
\wdo=\calD-\calL-\calU
\label{decomp_lu}
\ee
into block diagonal, block lower-triangular and block upper-triangular
parts with respect to position space, the SSOR preconditioner is defined
in terms of these matrices and a non-zero relaxation parameter $\omss$
(which serves to reduce the iteration number) through the choice
\be
\cl{V}_1=
\left(\frac{1}{\omss}\,\calD-\calL\right)
\left(\frac{1}{\omss}\,\calD\right)^{-1}\,,\quad
\cl{V}_2=
\frac{1}{\omss}\,\calD-\calU\,.
\label{v1v2}
\ee
Now it is advantageous to exploit in (\ref{ssor}) the so-called
`Eisenstat trick' \cite{solver:eisenstat}, embodied in the identity
\bea
&   &\,\,\,
\frac{1}{\omss}\,\calD\left(\frac{1}{\omss}\,\calD-\calL\right)^{-1}
\left(\calD-\calL-\calU\right)
\left(\frac{1}{\omss}\,\calD-\calU\right)^{-1}\nonumber\\
& = &\,\,\,
\left({\bf{1}}-\omss\calL\calD^{-1}\right)^{-1}\left\{
{\bf{1}}+\left(\omss-2\right)
\left({\bf{1}}-\omss\calU\calD^{-1}\right)^{-1}\right\}\nonumber\\
&   &\,\,\,
+\left({\bf{1}}-\omss\calU\calD^{-1}\right)^{-1}\,, 
\label{estat}
\eea
to save on computational costs:
namely, it implies that any matrix-vector product with the preconditioned
matrix $\cl{V}_1^{-1}\wdo\cl{V}_2^{-1}$ essentially gives rise to a
backward substitution and a subsequent forward substitution process,
corresponding to the application of the (non-block) upper and lower
triangular matrices ${\bf{1}}-\omss\calU\calD^{-1}$ and 
${\bf{1}}-\omss\calL\calD^{-1}$, respectively.
These relations reflect that SSOR splits any application of $\wdo$, which
usually enters the inversion procedure associated with (\ref{syst}), into
an equivalent but much simpler sequence of arithmetic operations with the
set of matrices $\{\calD,\calL,\calU\}$.

Then, combining the BiCGStab algorithm with SSOR preconditioning, an 
iterative numerical solution of the system in eq.~(\ref{syst}) up to a
given precision $\epsilon$ is obtained by the prescription\footnote{
The algorithm can straightforwardly formulated for other solvers, like
MR for instance. 
Note, however, that in the chiral quark mass regime MR is generally of
worse performance.}
(preconditioned quantities carry a tilde):
\bean
\mbox{\bf initialization with guess $X_0$ :}
& \quad &\\\\
R_0=Y-\wdo X_0
& \quad &\\
(\fat{1}-\omss\calL\calD^{-1})\tilde{R}_0=R_0
& \quad &\\\\
\tilde{\Rhat}\equiv\tilde{R}_0
& \quad &
\rho_0\equiv 1\\
\tilde{V}_0=\tilde{P}_0\equiv 0
& \quad &
\alpha_0=\omega_0\equiv 1\\\\
\mbox{\bf $k$-th iteration ($k\geq1$) :}
& \quad &\\
& \quad &
\rho_k=(\tilde{\Rhat},\tilde{R}_{k-1})\\\\
& \quad &
\beta=\frac{\rho_k\alpha_{k-1}}{\rho_{k-1}\omega_{k-1}}\\\\
\tilde{P}_k=
\tilde{R}_{k-1}+\beta \tilde{P}_{k-1}
-\beta\omega_{k-1}\tilde{V}_{k-1}
& \quad &\\\\
(\fat{1}-\omss\calU\calD^{-1})W=\tilde{P}_k\,,
\quad Z'=\tilde{P}_k+(\omss-2)W
& \quad &\\
(\fat{1}-\omss\calL\calD^{-1})Z=Z'\,,
\quad\tilde{V}_k=W+Z
& \quad &\\\\
& \quad &
\alpha_k=\frac{\rho_k}{(\tilde{\Rhat},\tilde{V}_k)}\\\\
\tilde{S}=\tilde{R}_{k-1}-\alpha_k\tilde{V}_k
& \quad &\\\\
(\fat{1}-\omss\calU\calD^{-1})U=\tilde{S}\,,
\quad Z'=\tilde{S}+(\omss-2)U
& \quad &\\
(\fat{1}-\omss\calL\calD^{-1})Z=Z'\,,
\quad\tilde{T}=U+Z
& \quad &\\\\
& \quad &
\omega_k=\frac{(\tilde{T},\tilde{S})}{(\tilde{T},\tilde{T})}\\\\
\tilde{X}_k=\tilde{X}_{k-1}+\omega_k U+\alpha_k W
& \quad &\\\\
\tilde{R}_k=\tilde{S}-\omega_k\tilde{T}
& \quad &
\eean
The stopping criterion to be imposed on the preconditioned residual is
\be
\frac{(\tilde{R}_k,\tilde{R}_k)}{(\tilde{X}_k,\tilde{X}_k)}
\le\epsilon^2\,.
\label{stop}
\ee
Some comments are in order.
\begin{itemize}
\item Taking the forward substitution as example, one calculates
      recursively in terms of the (block) components $Z_i$, $Z_i'$,
      $\calD_{ij}$ and $\calL_{ij}$ of $Z$, $Z'$, $\calD$ and $\calL$:
      \be
      \forall\,i:\quad
      Z_i=Z_i'+\sum_{j=1}^{i-1}\calL_{ij}H_j\,,\quad
      H_j=\omss\left(\calD^{-1}\right)_{jj}Z_j\,\Big|_{\,j\le i-1}\,.
      \label{fsubst}	
      \ee
      I.e.~thanks to the triangularity of $\calL$ (and $\calU$) it can be
      done economically without involving a plain matrix multiplication,
      and owing to the sparsity of $\wdo$ only a few $j$ actually
      contribute to the sum.
      Provided that the inverses $\left(\calD^{-1}\right)_{ii}$ are
      pre-computed, the backward and forward solves together are exactly
      as expensive as one application of the whole matrix $\wdo$ plus one
      additional $\calD$ multiplication.
\item In contrast to the unimproved case (where $X_k=\tilde{X}_k$ follows
      immediately), the solution of the original system $\wdo X=Y$ is now
      $X_k=\omss^{-1}\calD^{-1}\tilde{X}_k$.
\item Choose $X_0=0=\tilde{X}_0$ as initial guess for the solution to
      avoid an application of $\calD$, as part of $\wdo$, at the 
      beginning, which yields $R_0=Y-\wdo X_0=Y$.
      If $\calD$ is not needed elsewhere, this might be favourable in
      view of possible memory limitations of the hardware (e.g.~setting an
      upper bound on the accessible lattice volumes), because we made the
      experience that savings in iteration number when using an available
      solution of a foregoing inversion as initial guess are generically
      negligible.
\item Since in order to save on computational cost the stopping criterion
      (in our runs $\epsilon^2=10^{-12},10^{-13}$) is conveniently based
      on $\tilde{R}_k$, one might want to re-compute $R_k$ at the end
      to test for convergence again and eventually --- if the solution is
      not yet accurate enough --- to continue the iteration with a stronger
      stopping criterion imposed on $\tilde{R}_k$.
\item In the SSOR scheme the minimal number of vectors to be stored at
      each iteration $k$ is 9, if $\tilde{R}_{k-1}$ and $\tilde{S}$ share
      the same memory location, and if the source vector $Y$ is
      overwritten by the iterative solutions $\tilde{X}_k$.
\item Due to the overall Dirac and colour structure of the fermion field
      variables at every lattice point, $\wdo$ is genuinely partitioned
      into blocks of dimension $12\times12$.  
      The dependence of the speed of the SSOR preconditioner on the 
      diagonal (sub-)block size of $\calD_{ij}$ was not studied.
\end{itemize}
\subsection{Implementation for the Schr\"odinger functional}
\label{ImplSF}
For the implementation of SSOR preconditioning on a parallel computer, 
the ordering of the lattice points plays a key r\^{o}le\footnote{
We remark that traditional e/o preconditioning can be interpreted as SSOR
preconditioning of the even-odd ordered system \cite{ssor:wils_wupp}.}.
It determines the shape of $\wdo$ via the pattern of its non-zero entries
and thereby the degree of parallelism and the efficiency of the
preconditioner. 
Via adapting a locally-lexicographic scheme we closely follow
refs.~\cite{ssor:wils_wupp,ssor:impr_wupp}, where different orderings and
their consequences on the parallelization have been discussed.
Hence we only describe the salient issues characteristic for the SF
approach.

Assume that a given space-time lattice is matched on the, say, 
three-di\-men\-sion\-al grid of processing nodes of a parallel computer.
Then each node occupies a local lattice, where three of its extensions are
ratios of the total lattice sizes in the respective directions and the
corresponding numbers of processors.
The locally-lexicographic ordering (`colouring') is ensured by an
alphabetic ordering of the lattice sites belonging to these local lattices.
Associating a colour with each fixed position within the local lattices,
the original lattice is divided into groups, whose members couple to
sublattice points of different colours only.
In this way it becomes obvious that the forward and backward substitutions,
e.g.~to get $Z_i$ as in eq.~(\ref{fsubst}), can be handled in parallel
within the coloured groups, since for all positions of a given colour only
their predecessors --- in the lexicographic sense --- enter the necessary
computations\footnote{
So the general strategy would be to maximize the number of coloured groups,
while maintaining its strengths in accordance with the desired
parallelization.}.
Because among these lexicographically preceding sites, lattice points
living on the neighbouring processors contribute too, the quantities $H_i$
of (\ref{fsubst}) will have to be communicated from those processing nodes
to the site in question.

\begin{figure}[htb]
\begin{center}
\vspace{0.25cm}
\epsfig{file=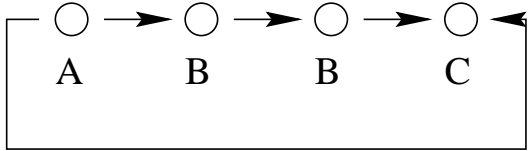,width=7.0cm}
\caption[t_param]{\label{1dPlot} \sl
 	          A one-dimensional example. 
                  The direction of the arrows shows the data flow:
                  sites of kind {\bf B} need information from site 
                  {\bf A}, and so forth.}
\vspace{0.25cm}
\end{center}
\end{figure}

\begin{figure}[htb]
\begin{center}
\epsfig{file=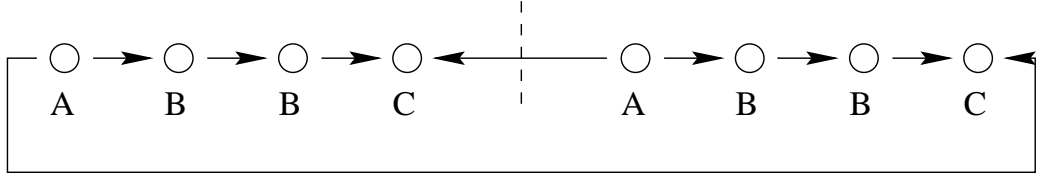,width=13.75cm}
\caption[t_param]{\label{1dpPlot} \sl
	          A one-dimensional parallel example.
                  As in the non-parallel case, figure~\ref{1dPlot}, the
                  arrows represent the data flow involved.}
\vspace{0.25cm}
\end{center}
\end{figure}

At this point we have to note that there is a crucial difference between a
situation with ordinary (anti-)periodic boundary conditions in all four
space-time directions and our SF setup with Dirichlet boundary conditions
in time.
To see this difference in more detail, let us resort to an example in one
dimension. 
The four points in figure~\ref{1dPlot} define a one-dimensional lattice
with periodic boundary conditions, and one can think of a `coupling matrix'
between the sites of mutual dependence.
%
%
Since only nearest neighbour sites are coupled, we immediately see that
in applying eq.~(\ref{fsubst}) one has to distinguish between three cases.
(We restrict to the forward solve, because it is straightforward to work
out the necessary changes for the backward solve.)
\begin{itemize}
\item[\bf{A}]
      The point is on the `left', i.e.~it has coordinate $x=1$: 
      (\ref{fsubst}) simply becomes
      \be
      Z_1=Z_1'\,.
      \ee
\item[\bf{B}]
      The point lies within the `bulk', i.e.~it has coordinate $x=2,3$:
      (\ref{fsubst}) becomes
      \be
      Z_x=Z_x'+\calL_{x,x-1}H_{x-1}\,,\quad x=2,3\,.
      \ee
\item[\bf{C}]
      The point is on the `right', i.e.~it has coordinate $x=4$:
      then (\ref{fsubst}) becomes
      \be
      Z_4=Z_4'+\calL_{4,3}H_{3}+\calL_{4,1}H_{1}\,.
      \ee
\end{itemize}
It is clear that we have those three cases for each dimension, in which
either periodic or antiperiodic boundary conditions are prescribed.
For a four-dimensional lattice this leads to $3^4=81$ different cases to be
handled, and it is natural to implement the algorithm with a
{\tt do}--loop over all the lattice points and some {\tt if}--statements to
discriminate between the 81 cases.
The \emph{parallel} version of the algorithm just described needs minor
modifications: looking at figure~\ref{1dpPlot}, only the case {\bf C} has
to be replaced with
\be
Z_4=Z_4'+\calL_{4,3}H_{3}+\calL_{4,1}H_{1}^{({\rm next\,processor})}\,,
\ee
where now the coordinate index has a local meaning, labelling the sites
inside the sublattice residing on a given processor, and also the
processor's mesh is assumed to have periodic boundary conditions in the
sense that the processor `next' to the last one in each direction
(rightmost in figure~\ref{1dpPlot}) is to be identified with the first one
in the same direction (leftmost in figure~\ref{1dpPlot}).
%
%

Our numerical simulations were done on the 8 -- 512 nodes APE-100
massively parallel computers with cubic topology and nearest neighbour
communication, built up of an array of elementary processing boards with
$2\times2\times2$ nodes \cite{computer:APE,computer:APE100}.
They possess a SIMD (single-instruction multiple-data) architecture and are
either suited to distribute a single, typically large, lattice over the
whole machine or to simulate in parallel several independent copies of a
lattice on subsets of the machine in the case of smaller volumes.

With such a cubic topology it suggests itself to keep the whole time extent
of the lattice within the processors, and to only split the spacelike
volume into sublattice fractions with respect to the three-dimensional
processor grid.
If \emph{SF boundary conditions} are adopted, one can make sure that,
as far as the time direction with its first and last timeslices fixed to
the boundary values is concerned, a site belongs always to the `bulk'.
This lowers the number of cases to be distinguished in order to implement
the forward and backward solves to $3^3=27$.
So it becomes rather near at hand to encode these cases explicitly in a
sensible arrangement of nested {\tt do}--loops alone, i.e.~via an outer
loop over the full time coordinate and inner ones over the coordinates of
the local space directions on every processor to exhaust the total lattice
volume; 
thereby the usage of any {\tt if}--statements is completely avoided.
Here arises the significant advantage of our implementation:
the latter type of statements cause --- especially so on the APE-100
machines --- a drastic deterioration of the performance by breaking the
so-called `optimization blocks' recognized by the compiler.
In practice, the contents of the registers is lost each time a branching
statement like {\tt if}, {\tt do}, {\tt call subroutine}, $\ldots\,$ is
encountered, because this amounts to a break in the memory pipeline
pre-loaded before.
Writing out the 27 distinct cases explicitly, however, reduces the impact
of the largest part of such statements, so that finally we are able to
arrive at a substantial speed-up of the code\footnote{
Of course one can imagine to write down analogously the 81 different
cases for the familiar periodic situation \cite{ssor:wils_wupp}, but then
the size of the executable file might easily exceed the integer and/or
program memory limits of the machine.}.
\subsection{Performance tests}
\label{Ptests}
\begin{figure}[htb]
\begin{center}
\vspace{-2.0cm}
\epsfig{file=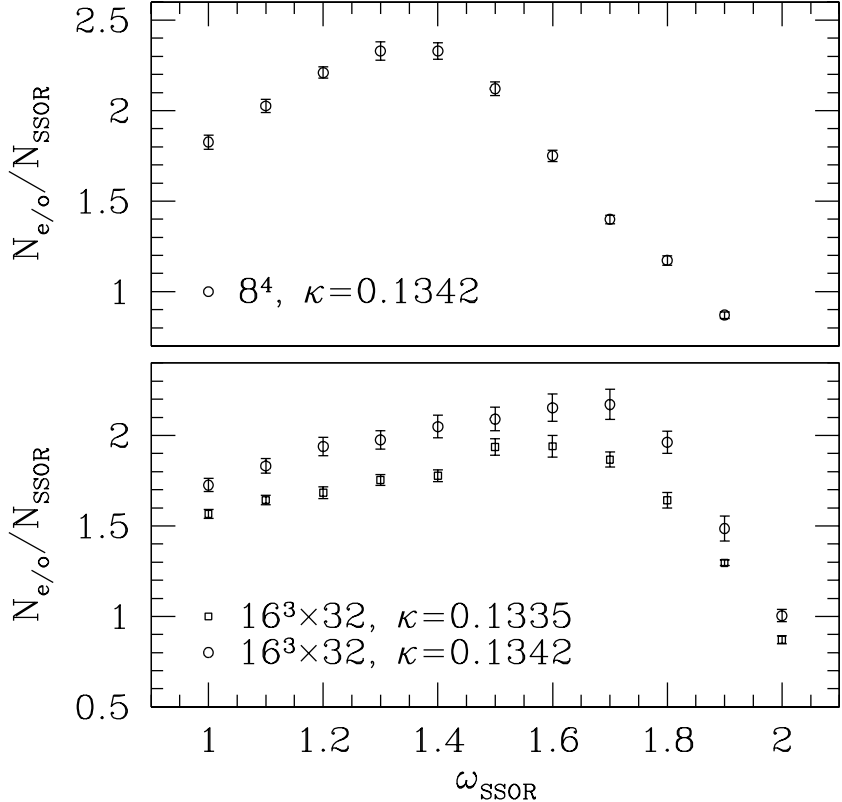,width=15.0cm}
\vspace{-1.5cm}
\caption[t_param]{\label{ItsPlot} \sl
                  Improvement factor in the number of BiCGStab iterations,
                  when simulating with SSOR and different choices for
                  $\omss$ instead of e/o preconditioning.
                  The lattice sizes are $8^4$ and $16^3\times32$, with
                  parameters $\beta=6.0$ and $\kappa=0.1335,0.1342$.}
\vspace{0.25cm}
\end{center}
\end{figure}

After realizing the implementation outlined above, we first have
investigated the influence of the relaxation parameter $\omss$ on the
numerical solution procedure for the linear system in eq.~(\ref{syst}).
It is quantified through the ratio of the numbers of iterations to solve
the system with the e/o preconditioned matrix, $\Neo$, and with the SSOR
preconditioned one, $\Nss$, under otherwise identical conditions.
We show these ratios in dependence on $\omss$, averaged over a set of 
$\Or(10)$ propagator computations occurring in quenched simulations, for
two lattice sizes with SF boundary conditions in figure~\ref{ItsPlot}.
%
%
Upon distributing these lattices over the meshes of nodes of the APEs, the
lattice extensions per node were $4^3\times8$ and $2^3\times32$.
The course of $\Neo/\Nss$ confirms a conclusion of the Wuppertal group
\cite{ssor:impr_wupp} that between $\omss\simeq1.3$ and $\omss\simeq1.6$
the gain in the number of iterations needed for the fermion matrix
inversions reaches its maximum: the corresponding improvement factor is
around or even above 2, while the tendency for a further increase of
$\Neo/\Nss$ towards the chiral limit (i.e.~larger $\kappa$ and smaller
quark mass) is also seen.
Above $\omss\simeq1.6-1.8$ this ratio drops rapidly and the gain gets
lost; therefore, we take over $\omss=1.4$ to be considered as an `optimal'
compromise irrespective of the definite values for lattice sizes and/or
parameters.

Now we pass to the central question, whether the gain in the number of
iterations also translates into a visible CPU time gain.
Of course, as already pointed out previously, this will depend on the
hardware architecture of the machine in use as well as on the individual
implementation.
At the peak in the upper diagram of figure~\ref{ItsPlot}, for instance,
a total saving of 1.5 in the spent simulation time can be attained.
In table~\ref{GainTab} we collect the approximate performance gain factors
in units of iteration number and net CPU time, which were found in the
situation of a realistic (quenched) QCD simulation in physically large
volumes.
%
\begin{table}[htb]
\begin{center}
\vspace{0.5cm}
\caption[t_param]{\label{GainTab} \sl
                  Approximate gain factors of SSOR over e/o in iteration
                  number and net CPU time}
\vspace{0.5cm}
\begin{tabular}{|cc|cc|}
\hline
  sublattice/node & subvolume/node & iteration gain & performance gain \\
\hline 
  $2^3\times32$        & $8\cdot32$  & 2.0 & 1.4 \\ 
  $2^2\times4\times32$ & $16\cdot32$ & 2.1 & 1.4 \\ 
  $2\times4^2\times32$ & $32\cdot32$ & 2.2 & 1.5 \\ 
\hline
\end{tabular}
\vspace{0.5cm}
\end{center}
\end{table}
%
\begin{figure}[htb]
\begin{center}
\vspace{-2.0cm}
\epsfig{file=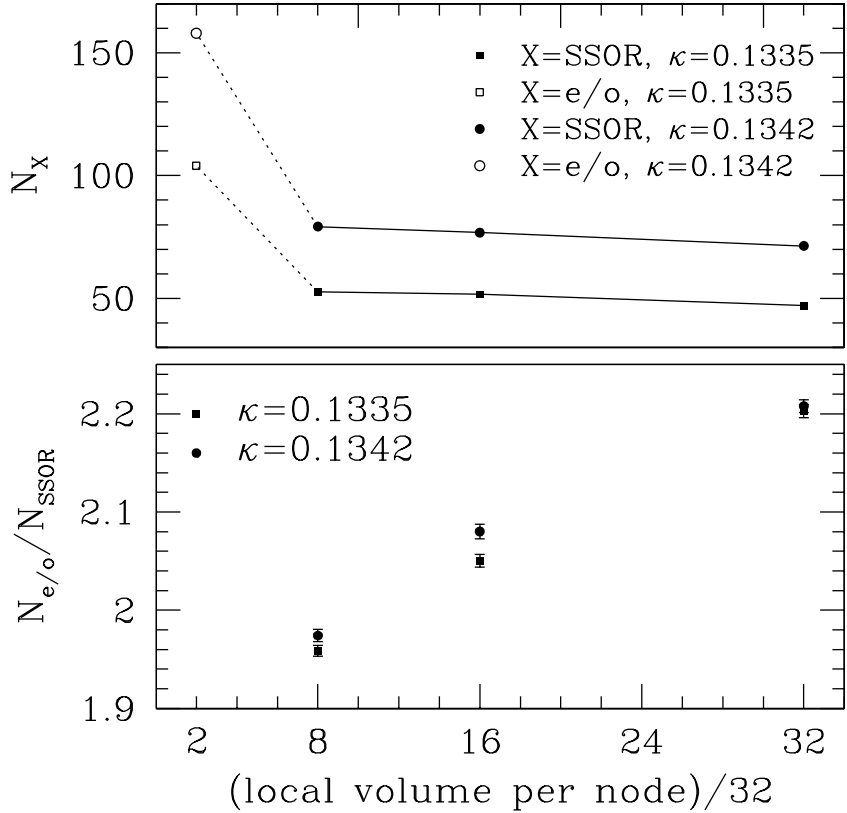,width=15.0cm}
\vspace{-1.5cm}
\caption[t_param]{\label{GainPlot} \sl
                  Upper part:
                  BiCGStab iteration number in dependence of the local
                  lattice volume per node after distributing a
                  $16^3\times32$ lattice on different three-dimensional
                  meshes of processing nodes.
                  Note that the common time extent of the local lattices,
                  $T=32$, has been divided out.
                  The e/o iteration numbers are included for comparison,
                  and the lines are only meant to guide the eye. 
                  Lower part:
                  Associated improvement factors in the number of
                  iterations.}
\vspace{0.5cm}
\end{center}
\end{figure}
%
Here we applied SSOR and e/o preconditioning for a set of 300 fermion
matrix inversions on thermalized $16^3\times32$ configurations at
$\beta=6.0$ and $\kappa=0.1335,0.1342$, where the relaxation parameter was
set to $\omss=1.4$ throughout and the pseudoscalar mass at those couplings
is $a\mps=0.388,0.300$ \cite{msbar:pap2,msbar:pap3}.
Moreover, we examined the dependence of the SSOR preconditioner on the
fractional grid size per node treated by a single processor.
As illustrated by the numbers in table~\ref{GainTab} and in
figure~\ref{GainPlot}, the iteration number ratio and thus the
preconditioning efficiency slightly increases with growing volumes of the
different local subgrids, $2^3\times32$, $2^2\times4\times32$ and
$2\times4^2\times32$, if the $16^3\times32$ lattice is spread over the
512, 256 and 128 processors of the available machines, respectively.
This complies with the heuristic expectation that an enlarged number of
coloured groups, i.e.~sets of points at the same fixed position within the
local sublattices (according to their locally-lexicographic ordering),
entails a measurable iteration gain in the inverter, whereas the
performance stays nearly unchanged owing to less parallelism and hence a
small accompanying inter-processor communication overhead.
Because of the equivalence of e/o preconditioning to a colouring into two
groups assigned to the even and odd sublattice, one can accommodate the e/o
iteration number in the upper part of figure~\ref{GainPlot} too.
Additionally, the points in both diagrams indicate once more the even
better behaviour of BiCGStab-SSOR in the range of lighter quark masses.

The foregoing observations are supported by the large scale simulations
in the strange quark mass region underlying the extraction of hadron masses
and matrix elements in quenched QCD with the SF reported in
refs.~\cite{msbar:pap2,msbar:pap3}.
There, at $\beta=6.1,6.2$ on $24^3\times T$ lattices (with $3^3\times T$
sublattices per node and $T=40,48$) and at $\beta=6.45$ on a $32^3\times64$
lattice (with $4^2\times8\times64$ sublattice per node), SSOR enabled to
save against e/o always CPU time factors of around $1.5-1.6$.

Altogether these results clearly reveal that the replacement of e/o by SSOR
preconditioning to solve the $\Or(a)$ improved Wilson-Dirac equation in the 
SF scheme pays off in real simulation costs.
In contrast to what the authors in \cite{ssor:impr_wupp} obtain, for the
same local diagonal block size of $\wdo$ and computer class, from QCD
simulations including the clover term with ordinary boundary conditions,
the SF type of boundary conditions allow for an efficient implementation of
SSOR preconditioning also on massively parallel machines with an
architecture equal or similar to that of the APE-Quadrics systems.

%% file: sect3.tex
\section{Conclusion}
\label{Concl}
We have demonstrated in numerical simulations of quenched lattice QCD with
the Sheikholeslami-Wohlert quark action that the increase of performance
between even-odd and SSOR preconditioning in a parallel implementation can
be a factor $\sim$ 1.5, when Schr\"odinger functional boundary conditions
are employed.

Opposed to the more standard situation with periodic boundary conditions
in all directions studied in ref.~\cite{ssor:impr_wupp}, the gain factors
in real time consumption come out to be significantly better here.
The main reason for this originates in the lower number of cases
(27 versus 81) to be distinguished explicitly, when the contributions to a
given site among the locally-lexicographically ordered points of the local
sublattices residing on the processors have to be collected: 
the avoidance of any conditional statements in the solver routines evades
unwanted breaks in the data flow within the registers of the computer, 
which then directly translates into a considerable gain in units of CPU
time.
We have to emphasize that this inherent sensitivity to pipeline
optimization might be --- at least partly --- a special feature of the 
APE-100 environment.
Nevertheless, since some conclusions drawn from the investigations in
\cite{ssor:impr_wupp_lat97,ssor:impr_wupp,ssor:impr_wupp_lat98} refer to
the identical particular machines, our findings should be of interest in
the same context and can be compared with the results stated there.

As the Schr\"odinger functional formulation of QCD is physically already
well accepted to be a viable framework to address many problems in the
non-perturbative low energy regime of the theory
\cite{schlad:rainer,msbar:pap3}, the observed evidence for a performance
gain of SSOR (together with BiCGStab as the inverter) for the $\Or(a)$
improved Wilson-Dirac operator in actual run time --- also on a parallel
computer --- constitutes a further benefit of this scheme.
Therefore, the feasibility of an efficient implementation of this
preconditioner does not only provide an important algorithmic information
by itself, but even more is also promising for any kind of future
applications of the Schr\"odinger functional in lattice QCD.
\subsection*{Acknowledgements}
This work is part of the ALPHA collaboration research programme.
We thank DESY for allocating computer time on the APE-Quadrics computers
at DESY Zeuthen to this project and the staff of the computer centre at
Zeuthen for their support.
We are grateful to Rainer Sommer for discussions, useful suggestions and a
critical reading of the manuscript.
J.\ H.\ also thanks Burkhard Bunk and Andreas Hoferichter for discussions.